\documentclass{emulateapj}
\usepackage{apjfonts}
\lefthead{CHUNG ET AL.} \righthead{MICROLENSING CENTRAL CAUSTICS}

\newcommand{\zetavec}{\mbox{\boldmath $\zeta$}}

\newcommand{\thetae}{\theta_{\rm E}}

\begin{document}
\title{Properties of Central Caustics in Planetary Microlensing}

\author{ Sun-Ju Chung\altaffilmark{1},
Cheongho Han\altaffilmark{1,9},
Byeong-Gon Park\altaffilmark{2},
Doeon Kim\altaffilmark{1},
Sangjun Kang\altaffilmark{3},
Yoon-Hyun Ryu\altaffilmark{4},
Kang Min Kim\altaffilmark{2},
Young-Beom Jeon\altaffilmark{2},
Dong-Wook Lee\altaffilmark{5},
Kyongae Chang\altaffilmark{6}, 
Woo-Baik Lee\altaffilmark{7}, and
Yong Hee Kang\altaffilmark{8} }
\altaffiltext{1}{Department of Physics, Institute for Basic Science
Research, Chungbuk National University, Chongju 361-763, Korea}
\altaffiltext{2}{Bohyunsan Optical Astronomy Observatory, Korea Astronomy 
and Space Science Institute, Youngchon 770-820, Korea}
\altaffiltext{3}{School of Liberal Arts, Semyung University, Jechon 390-711, 
Korea}
\altaffiltext{4}{Department of Astronomy and Atmospheric Sciences, 
Kyungpook National University, Daegu 702-701, Korea}
\altaffiltext{5}{Astrophysical Research Center for the Structure and 
Evolution of the Cosmos (ARCSEC''), Sejong University, Seoul 143-747, 
Korea}
\altaffiltext{6}{Department of Computer and Applied Physics, Chongju 
University, Chongju 360-764}
\altaffiltext{7}{Korea Astronomy and Space Science Institute, Taejon, 
Korea 305-348}
\altaffiltext{8}{Department of Earth Science Education, Kyungpook National 
University, Daegu 702-701, Korea}
\altaffiltext{9}{corresponding author; cheongho@astroph.chungbuk.ac.kr}

\begin{abstract}
To maximize the number of planet detections, current microlensing 
follow-up observations are focusing on high-magnification events 
which have a higher chance of being perturbed by central caustics.  
In this paper, we investigate the properties of central caustics 
and the perturbations induced by them.  We derive analytic expressions 
of the location, size, and shape of the central caustic as a function 
of the star-planet separation, $s$, and the planet/star mass ratio, 
$q$, under the planetary perturbative approximation and compare the 
results with those based on numerical computations.  While it has 
been known that the size of the planetary caustic is $\propto \sqrt{q}$, 
we find from this work that the dependence of the size 
of the central caustic on $q$ is linear, i.e., $\propto q$, implying 
that the central caustic shrinks much more rapidly with the decrease 
of $q$ compared to the planetary caustic.   The central-caustic size 
depends also on the star-planet separation.  If the size of the 
caustic is defined as the separation between the two cusps on the 
star-planet axis (horizontal width), we find that the dependence of 
the central-caustic size on the separation is $\propto (s+s^{-1})$.
While the size of the central caustic depends both on $s$ and $q$, 
its shape defined as the vertical/horizontal width ratio, ${\cal R}_c$, 
is solely dependent on the planetary separation and we derive an 
analytic relation between ${\cal R}_c$ and $s$.  Due to the smaller 
size of the central caustic combined with much more rapid 
decrease of its size with the decrease of $q$, the effect of finite 
source size on the perturbation induced by the central caustic is 
much more severe than the effect on the perturbation induced by the 
planetary caustic.  As a result, we find that although giant planets 
with $q\gtrsim 10^{-3}$ can be detected from the planet search 
strategy of monitoring high-magnification events, detecting signals 
of Earth-mass planets with $q\sim 10^{-5}$ will be very difficult.  
Although the central caustics of a pair of planets with separations 
$s$ and $s^{-1}$ are identical up to the linear order, we find that 
the magnification patterns induced by the pair of the degenerate 
caustics of planets with $q\gtrsim 10^{-3}$ are different to the 
level of being noticed from observations with $\lesssim 2\%$ photometry.  
Considering that the majority of planets to be detected by the strategy 
of monitoring high-magnification events are giant planets, we predict 
that the $s\leftrightarrow s^{-1}$ degeneracy could be broken for a 
majority of planetary events from observations with good enough precision.
\end{abstract}

\keywords{planetary systems -- planets and satellites: general -- 
gravitatinal lensing}

\section{Introduction}

Various methods have been proposed to detect and characterize extrasolar
planets, including radial velocity technique \citep{mayor95, marcy96}, 
transit method \citep{struve52}, direct imaging \citep{angel94, stahl95}, 
pulsar timing analysis \citep{wolszczan92}, and microlensing \citep{mao91, 
gould92}.  See also the review of \citet{perryman00}.  The microlensing 
signal of a planetary companion to Galactic disk and bulge microlens 
stars is the short-duration perturbation to the smooth standard light 
curve of the primary-induced lensing event occurred on background star.
Compared to other methods, the decay of the planetary lensing signal 
with the decrease of the planet/star mass ratio is relatively slow, 
and thus the microlensing technique has an important advantage of being 
applicable to the detections of Earth-mass planets by using already 
existing instrument \citep{bennett96}.  The microlensing method also 
has a unique applicability to the detections of free-floating planets 
\citep{bennett02, han04, han05}.  In addition, the method is not 
restricted to planets of nearby stars and can be extended even to 
nearby galaxies \citep{baltz01}. Recently, two clear-cut microlensing 
detections of exoplanets were reported by \citet{bond04} and 
\citet{udalski05}.

Due to the rare and incidental chance of lensing events combined 
with the short duration of planet signals, planetary lensing searches 
require a special observational setup, where survey observations 
issue alerts of ongoing events in the early stage of lensing 
magnification \citep{soszynski01, bond01} and follow-up collaborations 
intensively monitor the alerted events \citep{bond02, park04, cassan04}.  
However, follow-up is generally done with small field-of-view instrument, 
and thus events are monitored sequentially.  As a result, only a handful 
number of events can be followed at any given time, limiting the number 
of planet detections.  To maximize the number of potential planet 
detections with a limited use of resources and time, \citet{griest98} 
proposed to focus on high-magnification events.  They pointed out that 
high-magnification events have a dramatically higher chance of being 
perturbed by planets due to the existence of a `central' caustic 
(see \S\ 2 for more detail).  By adopting this proposal, current 
follow-up experiments are giving high priority to these events 
\citep{albrow01, rattenbury02, bond02, abe04, jiang04}.  However, little has been studied 
about the characteristics of central caustics and the perturbations 
induced by them (central perturbations).

In this paper, we investigate the properties of central caustics and 
the perturbations induced by them.  The layout of the paper is as 
follows.  In \S\ 2, we describe basic physics of planetary lensing.  
In \S\ 3, we analytically investigate how the location, size, and 
shape of the central caustic vary depending on the mass ratio, $q$, 
and separation, $s$, of planets under the planetary perturbative 
approximation and compare the results with those based on numerical 
computations.  In \S\ 4, we then systematically inspect the patterns 
of central perturbations by constructing maps of magnification excess 
for planets with various values of $s$ and $q$.  From the constructed 
maps, we also examine the effect of finite source size and investigate 
the possible types of planets detectable for given source stars with 
various sizes.  We summarize the results and conclude in \S\ 5.

\section{Basics of Planetary Lensing}

The lensing mapping from the lens plane to the source plane of $N$-point 
masses with no external shear or convergence is described by the lens 
equation of 
\begin{equation}
\zeta = z - \sum_{j=1}^N {m_j/M \over \bar{z}-\bar{z}_{L,j}},
\label{eq2.1}
\end{equation}
where $\zeta=\xi + i\eta$, $z_{L,j}=x_{L,j}+iy_{L,j}$, and $z=x+iy$ 
are the complex notations of the source, lens, and image positions, 
respectively, $\bar{z}$ denotes the complex conjugate of $z$, $m_j$ 
are the masses of the individual lens components, and  $M=\sum_j m_j$ 
is the total mass of the lens system \citep{witt90}.  Here all angles 
are normalized to the Einstein ring radius $\thetae$ of the total mass 
of the system, i.e, 
\begin{equation}
\thetae={r_{\rm E} \over D_{\rm L}}=\left[ {4GM\over c^2} 
\left( {1\over D_{\rm L}} - {1\over D_{\rm S}}  \right)
\right]^{1/2},
\label{eq2.2}
\end{equation}  
where $D_{\rm L}$ and $D_{\rm S}$ are the distances to the lens and 
source, respectively.  For a single lens $(N=1)$, there exist two 
images with locations at $u_{I,\pm}=0.5[u\pm (u^2+4)^{1/2}]$ and 
magnifications of $A_\pm=0.5(A\pm 1)$, where $u\equiv |\zeta-z_{L}|$ 
is the separation between the lens and source.  Then, the total 
magnification corresponds to the sum of the magnifications of the 
individual images and it is related to $u$ by
\begin{equation}
A=A_{+} + A_{-}={(u^2+2) \over u(u^2+4)^{1/2} }.
\label{eq2.3}
\end{equation}

A planetary lensing is described by the formalism of a binary ($N=2$) 
lens with a very low-mass companion.  For a binary lens, there are 
three or five images and the number of images changes by two as the 
source crosses a caustic.  The caustics are important features of 
binary lensing and they represent the set of source positions at 
which the magnification of a point source becomes infinite.  The 
caustics of binary lensing form a single or multiple closed figures 
where each figure is composed of concave curves (fold caustics) that 
meet at cusps.  For a planetary case, there exist two sets of 
disconnected caustics: one `central caustic' located close to the 
host star and one or two `planetary caustics' depending on whether 
the planet lies outside or inside the Einstein ring.  Since the 
central caustic lies close to the host star, the perturbation induced 
by the central caustic occurs close to the peak of lensing light 
curves of high-magnification events.  For a star-planet system, it 
is known that the lensing behavior can be described by the perturbation 
approach due to the small planet/star mass ratio \citep{bozza99, 
asada02, an05}.  In this case, the lensing equation is expressed as 
\begin{equation}
\zeta = z - {1\over \bar{z}} - {q\over {\bar{z}-\bar{z}_p}},
\label{eq2.4}
\end{equation}
where the position of the star is chosen as the coordinate origin and 
$z_p$ represents the location of the planet.

\begin{figure}[tb]
\epsscale{0.95}
\plotone{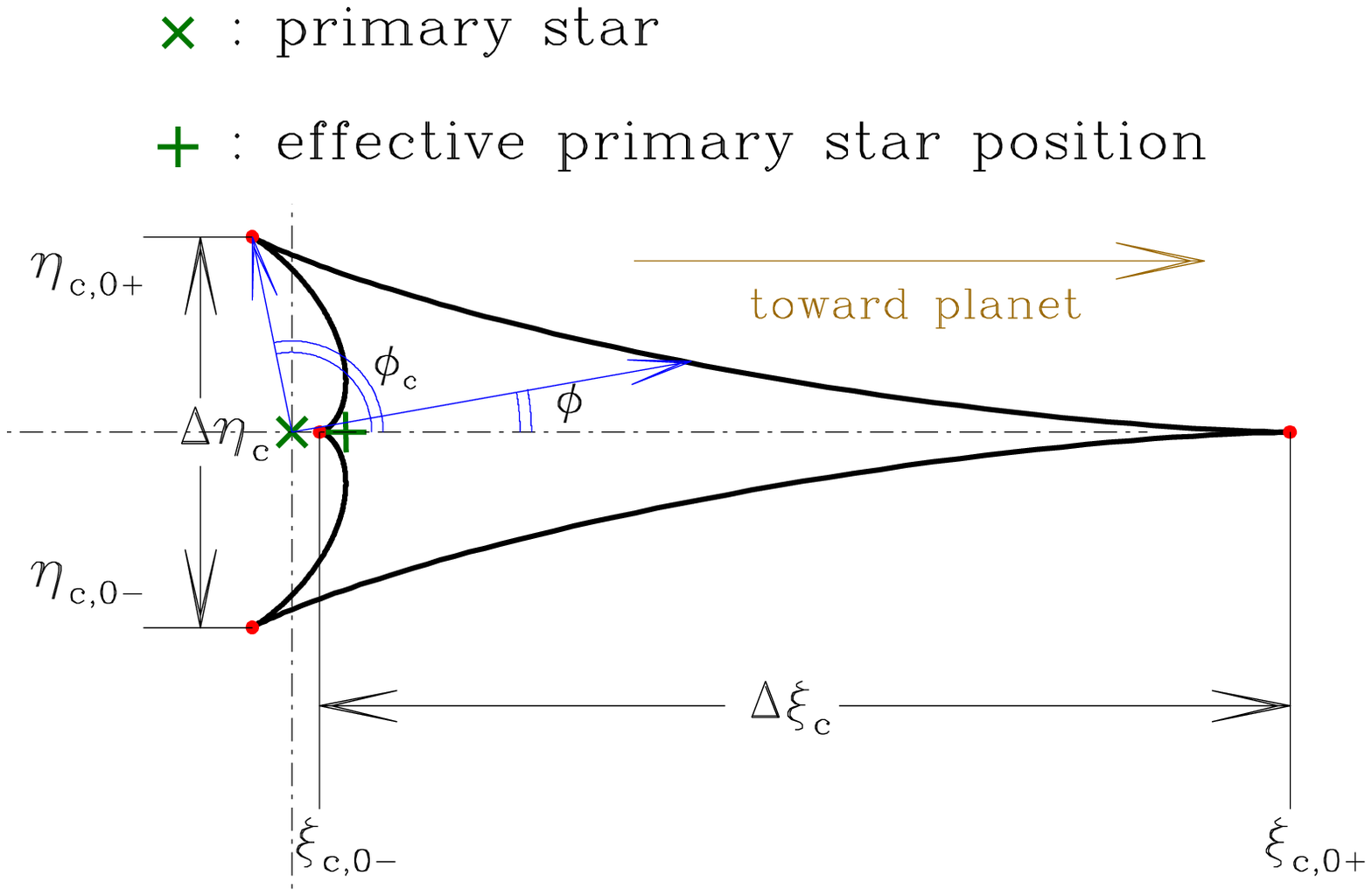}
\caption{\label{fig:one}
Geometry of a central caustic.
}\end{figure}

\begin{figure*}[t]
\epsscale{0.75}
\caption{\label{fig:two}
Variations of the shape and size of the central caustic produced by 
planets with various separations from and mass ratios relative to 
the primary star.  The left panels show the variations of the shape 
of the central caustic depending on the separation, while the right 
panels show the variation depending on the mass ratio.  In all cases, 
coordinates are centered at the {\it effective} position of the host 
star and the planets are located on the left side.  Note that the 
axis scales differ from one panel to another for better comparison of 
the caustic shapes.
}\end{figure*}

\section{Properties of Central Caustics}

Under the planetary perturbative 
approximation ($q\ll 1$ and $||z_p|-1|\gg q$), the location of the 
central caustic can be expressed in an analytic form of 
\begin{equation}
\zeta_c \simeq {q\over 2} e^{i\phi} 
\left[ 
{1\over (1-z_pe^{-i\phi})^2} +
{1\over (1-\bar{z}_p^{-1}e^{-i\phi})^2} - 1
\right]
\label{eq3.1}
\end{equation}
\citep{an05}, 
where the polar coordinates are centered at the position of the primary 
star and $\phi$ represents the polar angle (see Figure~\ref{fig:one}).  
With the notation for the star-planet separation of $s=|z_p|$, the 
caustic position is expressed in an explicit form of 
\begin{equation}
\xi_c \simeq { s+s^{-1}+2(\cos^3\phi-2\cos\phi) \over 
(s+s^{-1}-2\cos\phi)^2 } q,
\label{eq3.2}
\end{equation}
\begin{equation}
\eta_c \simeq -{2\sin^3\phi \over   (s+s^{-1}-2\cos\phi)^2 }q.
\label{eq3.3}
\end{equation}
The central caustic has an elongated asteroid shape with four cusps, 
where two of them are located on the star-planet axis and the other 
two are located off the axis.  Cusps of a central caustic occur when 
$d\zeta_c/d\phi=0$ (i.e., when $\phi=0$ and $\pi$ for the on-axis cusps 
and when $\phi=\phi_c$ and $2\pi-\phi_c$ for the off-axis cusps), and 
thus their locations are found to be 
\begin{equation}
\xi_{c,0\pm} \simeq \pm {q\over (1\pm s)(1\pm s^{-1})},
\label{eq3.4}
\end{equation}
\begin{equation}
\eta_{c,0\pm} \simeq \pm{2q |\sin^3\phi_c|\over (s+s^{-1}-2\cos\phi_c)^2},
\label{eq3.5}
\end{equation}
where 
\begin{equation}
\cos\phi_c = {3\over 4}\left( s+{1\over s}\right)
\left[ 1-\sqrt{ 1-{32\over 9} \left(s+{1\over s}\right)^{-2}  }\right].
\label{eq3.6}
\end{equation}
Then, the horizontal and vertical widths of the central caustic defined 
as the separations between the on- and off-axis cusps 
(see Figure~\ref{fig:one}) are respectively
\begin{equation}
\Delta\xi_c \simeq  
|\xi_{c,0+} - \xi_{c,0-}| =
{4q\over (s-s^{-1})^{2}},
\label{eq3.7}
\end{equation}
\begin{equation}
\Delta\eta_c \simeq 
|\eta_{c,0+} - \eta_{c,0-}| =
\Delta\xi_c {(s-s^{-1})^2|\sin^3\phi_c|
\over (s+s^{-1}-2\cos\phi_c)^2}.
\label{eq3.8}
\end{equation}

\begin{figure}[t]
\epsscale{0.95}
\plotone{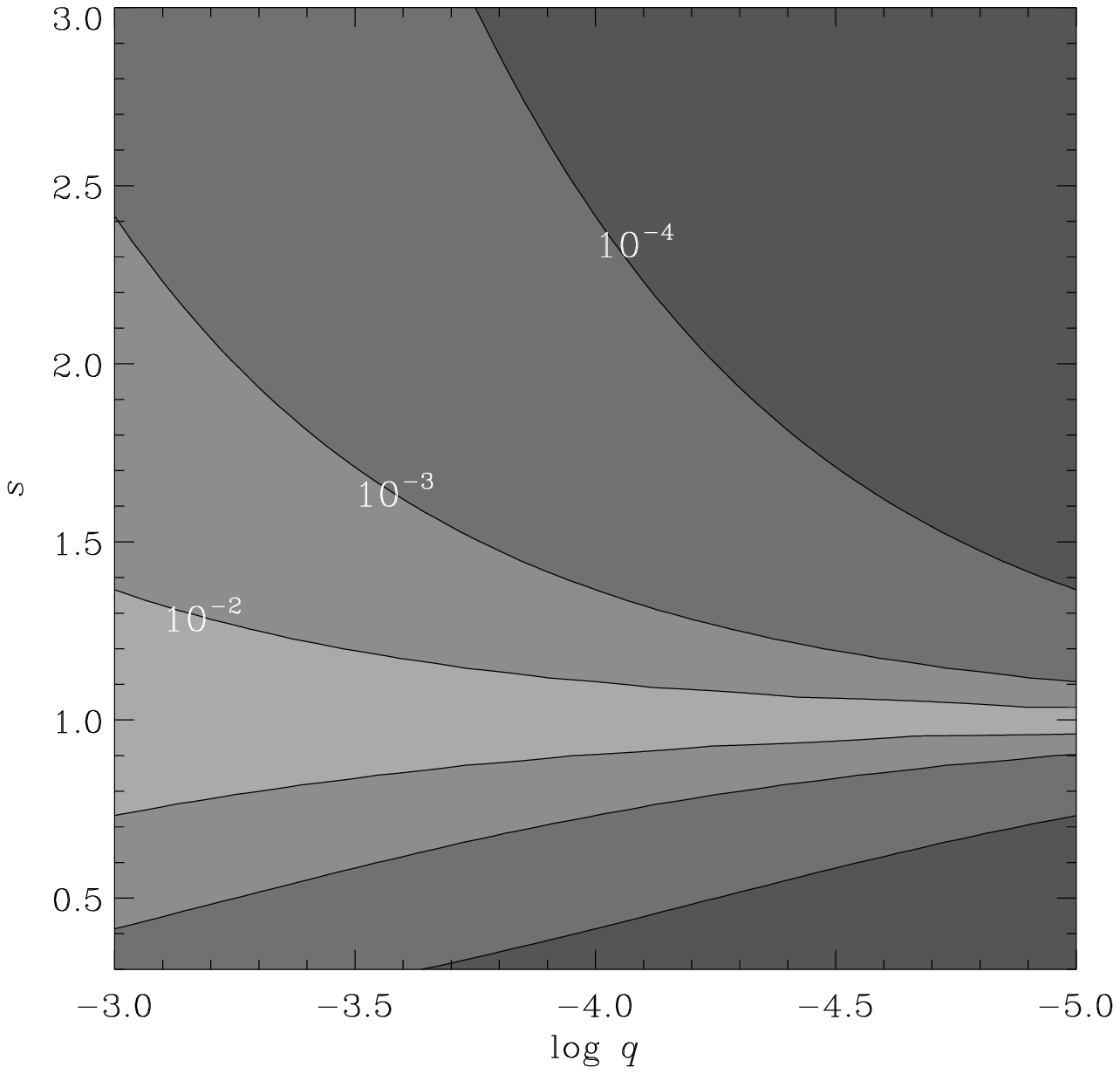}
\caption{\label{fig:three}
Size of the central caustic, as measured by the horizontal width, as 
a function of $s$ and $q$.
}\end{figure}

\begin{figure}[t]
\epsscale{0.95}
\plotone{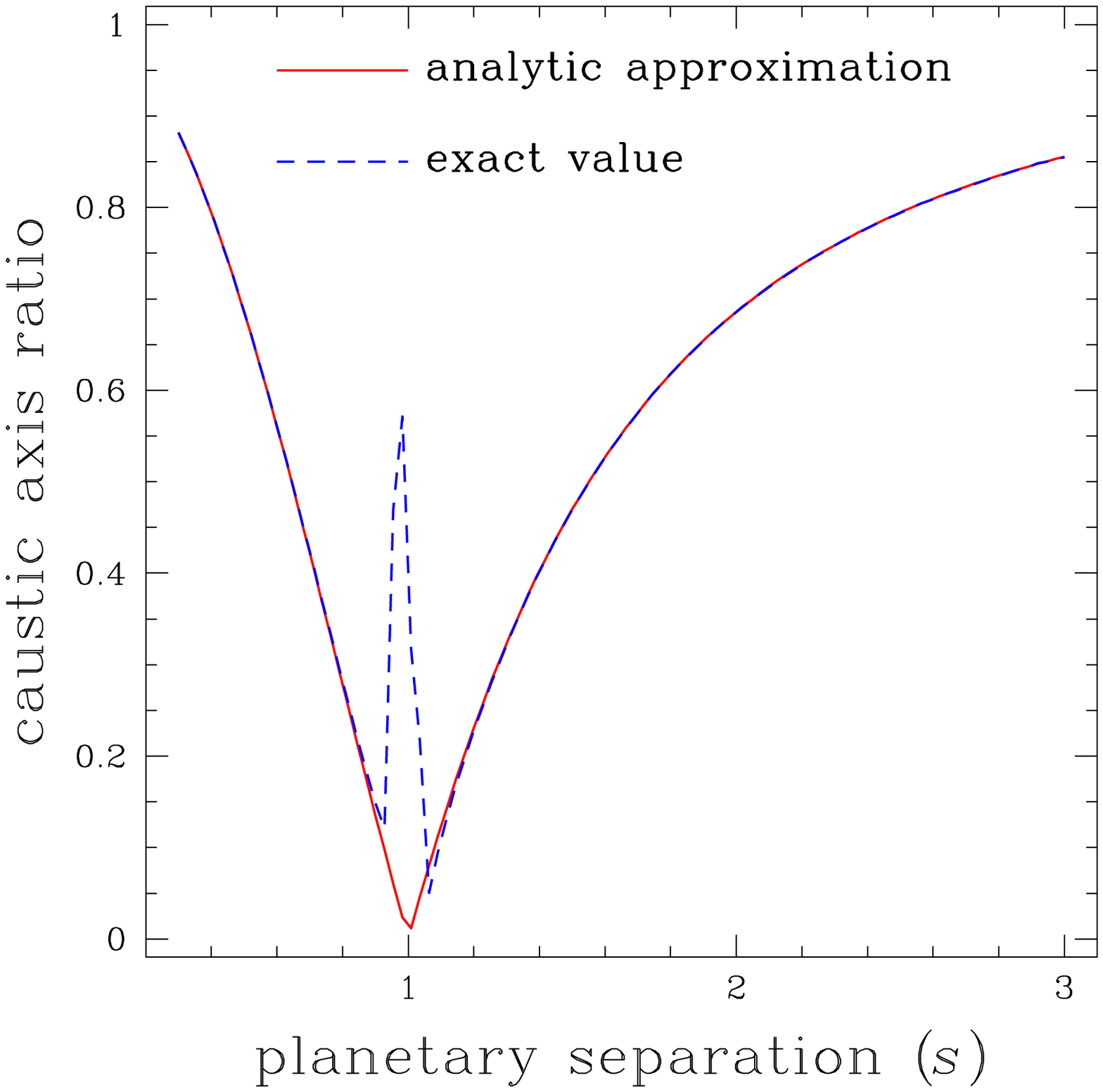}
\caption{\label{fig:four}
Dependence of the central-caustic shape, as measured by the ratio between 
horizontal and vertical widths, on the planet-star separation.  The 
computations are based on planets with $q=10^{-4}$.
}\end{figure}

\begin{figure*}[t]
\epsscale{0.95}
\caption{\label{fig:five}
Contour maps of magnification excess as a function of source position 
$(\xi,\eta)$ for planetary lens systems with various star-planet 
separations and planet/star mass ratios in the region around central 
caustics.  Contours are drawn at the levels of $\epsilon=\pm 2\%$ 
(thin curves) and $\pm 5\%$ (thick curves) and greyscale is used to 
represent positive (bright) and negative (dark) deviation regions.  
All lengths are scaled by the Einstein ring radius corresponding to 
the total mass of the lens system.  The coordinates are centered at 
the {\it effective} position of the host star.  In all cases, planets 
are located on the left side.  The maps are constructed for a 
{\it main-sequence} source star with a normalized radius of 
$\rho_\star=0.0018$, that corresponds to the case where the source 
with an absolute radius of $R_\star=1\ R_\odot$ and located at
 $D_{\rm S}=8\ {\rm kpc}$ is lensed by a lens with $m=0.3\ M_\odot$ 
and $D_{\rm L}=6\ {\rm kpc}$.
}\end{figure*}

\begin{figure*}[t]
\epsscale{0.9}
\caption{\label{fig:six}
Contour maps of magnification excess for the same planetary lens systems
as in Fig.~\ref{fig:five}, but for events associated with a {\it turn-off} 
source star with $\rho_\star=0.0054$.
}\end{figure*}

\begin{figure*}[t]
\epsscale{0.9}
\caption{\label{fig:seven}
Contour maps of magnification excess for the same planetary lens systems
as in Fig.~\ref{fig:five}, but for events associated with a {\it clump-giant} 
with $\rho_\star=0.0234$.
}\end{figure*}

In Figure~\ref{fig:two}, we present central caustics produced by planets 
with various separations and mass ratios.  From the figure and equations 
(\ref{eq3.1}) -- (\ref{eq3.8}), we find the following characteristics of 
the central caustic.

\begin{enumerate}

\item
The size of the central caustic depends on both the separation and mass 
ratio.  The dependence of the central-caustic size on the mass ratio is 
linear, i.e., $\propto q$, as shown in equations~(\ref{eq3.1}) -- 
(\ref{eq3.3}) and demonstrated in the right panels of Figure~\ref{fig:two}.  
By comparison, the size of the planetary caustic is proportional to 
$\sqrt{q}$.  This implies that the central caustic shrinks much more 
rapidly with the decrease of the planet mass compared to the planetary 
caustic.  If the caustic size is defined as the horizontal width, the 
size depends on the star-planet separation by $\Delta\xi_c\propto 
(s+s^{-1})^{-2}$.  Then, the caustic size becomes maximum when $s\sim 1$ 
and decreases with the increase of $|s-1|$.  In the limiting cases of 
a very wide-separation planet ($s\gg 1$) and a close-in planet ($s\ll 1$),
the dependencies are respectively
\begin{equation}
\Delta\xi_c \propto 
\cases{
s^{-2}  & for $s \gg 1$, \cr
s^2     & for $s \ll 1$.\cr
}
\label{eq3.9}
\end{equation}
In  Figure~\ref{fig:three}, we present the size of the central caustic 
as a function of $s$ and $q$.

\item
For a given mass ratio, a pair of central caustics with separations $s$ 
and $s^{-1}$ are identical up to linear order.  This can be seen from 
the pair of caustics with separations $s$ and $s^{-1}$ presented in 
Fig.~\ref{fig:two}.  Analytically, this can also be proved from 
equations~(\ref{eq3.1}) -- (\ref{eq3.3}) where the inversion of 
$s\leftrightarrow s^{-1}$ result in the same expression.

\item
Under the perturbative approximation,
the shape of the central caustic is solely dependent on the planet 
separation.  We quantify the caustic shape as the vertical/horizontal 
width ratio, ${\cal R}_c=\Delta\eta_c / \Delta\xi_c$, and present the 
variation of ${\cal R}_c$ as a function of $s$ in Figure~\ref{fig:four}.  
In the figure, we present two curves, where one is based on the 
perturbative approximation and the other is based on numerical computation.  
One finds that the width ratios based on the analytic and numerical 
computations match very well except the region around $s \sim 1$, 
where the perturbative approximation is not valid.  We find that the 
difference in this region is caused by the merge of the planetary and 
central caustics.\footnote{The planetary caustic produced by a planet 
with a separation $s$ from its parent star is located at ${\bf x}_{pc}
\sim {\bf s}-1/{\bf s}$. Therefore, as $s\rightarrow 1$, $x_{pc}
\rightarrow 0$, and thus the planetary caustic approaches the central 
caustic and eventually merge.}  One also finds that the width ratio of 
the central caustic rapidly increases with the decrease of $|s-1|$.
\end{enumerate}

\section{Central Caustic Perturbations}

Knowing now the characteristic of central caustics, we then investigate 
the pattern of planetary perturbations induced by central caustics.
For this purpose, we construct maps of magnification excess, which is  
defined as
\begin{equation}
\epsilon(\xi,\eta) = {A-A_0\over A_0},
\label{eq4.1}
\end{equation}
where $A$ is the exact magnification of the planetary lensing and $A_0$ 
is the single lensing magnification caused by the host star at its 
`effective' position.  Due to the additional deflection of light 
produced by the presence of the companion, it was known that the 
effective lensing position of a component of the binary lens system 
is shifted toward its companion \citep{distefano96, an02}.  The 
amount of the shift of a lens component `$i$' toward the other 
component `$j$' is 
\begin{equation}
\Delta x_{L,i\rightarrow j}\simeq {m_j/m_i \over (s+s^{-1})/
(\theta_{{\rm E},i}/\theta_{\rm E})} 
{\theta_{{\rm E},i}\over \theta_{\rm E}},
\label{eq4.2}
\end{equation} 
where $m_i$ and $m_j$ are the masses of the individual lens components
and $\theta_{{\rm E},i}$ is the Einstein ring radius corresponding to 
$m_i$ \citep{an02}.  For a planetary lens case, the shift of the host 
star (with a notation `$\star$') toward the planet (with a notation 
`$p$') is expressed as
\begin{equation}
\Delta x_{L,\star\rightarrow p}\simeq  {q \over (s+s^{-1})},
\label{eq4.3}
\end{equation}
because $\theta_{{\rm E},\star}\sim\theta_{\rm E}$ and 
$m_p/m_\star=q$.\footnote{We note that the term `$s^{-1}$' in 
equations~(\ref{eq4.2}) and (\ref{eq4.3}) was not included in the 
corresponding equations of \citet{distefano96} and \citet{an02}, where 
they treated only wide-separation binaries.  We introduce this term 
to keep the symmetry between the central caustics of planets with 
separations $s$ and $s^{-1}$.} Then, as either $|s-1|\rightarrow \infty$ 
(wide-separation planet) or $|s-1|\rightarrow 0$ (close-in planet),
the shift is $\Delta x_{L,\star\rightarrow p}\rightarrow 0$, and thus 
the effective lensing position of the primary star approaches its 
original position.

\begin{deluxetable}{lccc}
\tablecaption{Detectability of Planetary Signals \label{table1}}
\tablewidth{0pt}
\tablehead{
\multicolumn{1}{c}{source} &
\multicolumn{3}{c}{detectability} \\
\colhead{type} &
\colhead{$q=10^{-3}$} &
\colhead{$q=10^{-4}$} &
\colhead{$q=10^{-5}$} }
\startdata
main-sequence ($\rho_\star=0.0018$) & O   & O                & X \\
turn-off      ($\rho_\star=0.0054$) & O   & $\bigtriangleup$ & X \\
clump-giant   ($\rho_\star=0.0234$) & O   & X                & X \\
\enddata
\tablecomments{ 
The adopted source radius are $R_\star=1\ R_\odot$, $3\ R_\odot$, and 
$13\ R_\odot$ for the main-sequence, turn-off, and clump giant 
stars, respectively.  The normalized source radius $\rho_\star =
\theta_\star/\theta_{\rm E}$ of each source star is determined by 
choosing an Einstein ring radius of $\theta_{\rm E}=0.32$ mas that 
corresponds to the value of the most frequent Galactic bulge event 
with $m=0.3\ M_\odot$, $D_{\rm L}=6\ {\rm kpc}$, and $D_{\rm S}=8\ 
{\rm kpc}$.  The individual symbols for the detectability represent
possible ('O'), marginal ('$\bigtriangleup$'), and very difficult
('X'), respectively.
}
\end{deluxetable}

In Figure~\ref{fig:five} -- \ref{fig:seven}, we present the constructed 
magnification-excess maps for planets with various values of $s$ and $q$.  
Planetary signal is important for planets with separations within the so 
called `lensing zone' of $1/1.6\lesssim s \lesssim 1.6$ \citep{gould92}, 
and thus we plot maps for planets located within this range.  To see the 
effect of finite-source size, we construct three sets of maps with source 
stars of main-sequence (with a radius of $R_\star=1\ R_\odot$), turn-off 
(with $R_\star=3\ R_\odot$), and clump giant (with $R_\star= 13\ R_\odot$) 
stars and they are presented in Figure~\ref{fig:five}, \ref{fig:six}, 
and \ref{fig:seven}, respectively.  The magnification of a finite source 
with a surface brightness profile of $I(\zetavec)$ is computed by the 
intensity-weighted magnification averaged over the source star surface, 
i.e., 
\begin{equation}
A_{fs}(\zetavec)=
{\int_S I(\zetavec') A(\zetavec+\zetavec') d \zetavec'
\over 
\int_S I(\zetavec')d \zetavec'},
\label{eq4.4}
\end{equation}
where $A$ denotes the point-source magnification, $\zetavec$ is the 
vector position of the center of the source, $\zetavec'$ is the 
displacement vector of a point on the source star surface with respect 
to the source star's center, and the two-dimensional integral is over 
the source-star surface $S$.  The effect of finite source size is 
smearing out the detailed structures of the planetary lensing signals.  
The finite-source effect is determined by the {\it normalized} source 
radius $\rho_\star$, which represents the angular radius of a star, 
$\theta_\star$, in units of the Einstein ring radius, i.e.\ 
$\rho_\star=\theta_\star/ \theta_{\rm E}$.  We thus set $\rho_\star$ 
of the individual stars by choosing an Einstein radius of 
$\theta_{\rm E}=0.32\ {\rm mas}$ that corresponds to the value of 
the most frequent Galactic bulge event with $m=0.3\ M_\odot$, 
$D_{\rm L}=6\ {\rm kpc}$ and $D_{\rm S}=8\ {\rm kpc}$.  Then, the 
normalized source radius of a $R_\star=1\ R_\odot$ star corresponds 
to $\rho_\star=0.0018$.  For a source with uniform surface brightness, 
the computation can be reduced from a two-dimensional to a one-dimensional 
integral using the Generalized Stokes's theorem \citep{gould97, dominik98}.  
To accelerate computations, we thus assume that the source stars have 
uniform surface brightness.  However, we note that the effect of 
non-uniform surface brightness on the planetary lensing signal is not 
important.    For a planetary system with a primary star of 
$m=0.3\ M_\odot$, the mass ratios of the Jupiter-, Saturn-, Neptune-, 
and Earth-mass planets correspond to $q\sim 3\times 10^{-3}$, $10^{-3}$, 
$10^{-4}$, and $10^{-5}$, respectively.

From the excess maps, we find the following properties of central 
perturbations.
\begin{enumerate}
\item
First, we find that the effect of finite source size on central 
perturbations is much more severe than the effect on perturbations 
caused by planetary caustics.  This is because the central caustic 
is not only much smaller but also shrinks much more rapidly with 
the decrease of $q$ than the planetary caustic.  As a result, we find 
that although giant planets with $q\gtrsim 10^{-3}$ can be detected 
from the planet search strategy of monitoring high-magnification events, 
detecting signals of Earth-mass planets with $q\sim 10^{-5}$ would be 
very difficult.  It is estimated that the lower mass limit of detectable 
planets will be that of a Neptune-mass planet with $q\sim 10^{-4}$, but 
detecting this mass-range planets will be possible only for events 
associated with source stars smaller than turn-off stars. In 
Table~\ref{table1}, we summarize the possible types of planets detectable 
from events involved with various types of source stars.
\item
Second,
as expected by the close similarity between the pair of caustics with 
separations $s$ and $s^{-1}$, the pattern of perturbations induced by 
these two caustics are similar each other.  A good example is the 
pair of maps with $(s,q)=(1.2,10^{-4})$ and $(1/1.2,10^{-4})$ in 
Figure~\ref{fig:five}.  However, we find that as the planet mass 
increases, the difference between the perturbation patterns of the 
pair of planets increases.  We find that the difference is large enough 
to be noticed for planets with $q\gtrsim 10^{-3}$, c.f., the pair of 
maps of the planets with $(s,q)=(1.2,10^{-3})$ and $(1/1.2,10^{-3})$.  
Considering that the majority of planets to be detected by the 
strategy of monitoring high-magnification events are giant planets, we 
predict that the $s\leftrightarrow 1/s$ degeneracy could be broken for 
a majority of planetary events from observations with a $\lesssim 2\%$ 
photometric precision.
\end{enumerate}

\section{Summary and Conclusion}

We investigated the properties of central caustics and the perturbations 
induced by them.  Under the planetary perturbative approximation, we 
derived analytic expressions of the location, size, and shape of the 
central caustic as a function of the planet separation and mass ratio 
and compared the results with those based on numerical computations.  
While it has been known that the size of the planetary caustic is 
proportional to $\sqrt{q}$, we found from this work that the dependence 
of the central-caustic size on the mass ratio is linear.  As a result, 
the central caustic shrinks much more rapidly with the decrease of the 
planet mass compared to the planetary caustic.  Due to the large 
finite-source effect caused by the smaller size of the central caustic 
combined with much more rapid decrease of its size with the decrease 
of $q$, we predict that although giant planets with $q\gtrsim 10^{-3}$ 
can be detected from the planet search strategy of monitoring 
high-magnification events, detecting signals of Earth-mass planets 
with $q\sim 10^{-5}$ would be very difficult.  Although the central 
caustics of a pair of planets with separations $s$ and $s^{-1}$ are 
identical up to the linear order, we found that the difference between 
the magnification patterns induced by the pair of degenerate caustics 
of planets with $q\gtrsim 10^{-3}$ can be noticed at the level of 
$\sim 2\%$.  Because the majority of planets expected to be detected 
by the strategy of monitoring high-magnification events are giant 
planets, we predict that the $s\leftrightarrow s^{-1}$ degeneracy could 
be broken for a majority of planetary events from observations with 
good enough photometric precision.

\acknowledgments 
We would like to thank J.\ H. An for making helpful comments.  
Work by C.H. was supported by the Astrophysical Research Center for 
the Structure and Evolution of the Cosmos (ARCSEC") of Korea Science 
and Engineering Foundation (KOSEF) through Science Research Program 
(SRC) program.  B.-G.P. acknowledges the support from the grant of 
Korea Astronomy and Space Science Institute (KASI).

\end{document}